\documentclass[prd,nofootinbib,11pt]{revtex4}
\usepackage{amssymb}
\usepackage{mathrsfs}
\usepackage[utf8]{inputenc}
\usepackage[T1]{fontenc}
\usepackage[english]{babel}
\usepackage{hyperref}
\usepackage{amsmath}
\usepackage{amsfonts}

\def\beq{\begin{equation}}
\def\eeq{\end{equation}}

\begin{document}
\title{Constructive Gravity: Foundations and Applications\\
{\it Rapporteur article for the session on Constructive Gravity\\ 15th Marcel Grossmann Meeting 2018 in Rome}
}
\author{Frederic P. Schuller}
\email{E-mail: f.p.schuller@utwente.nl}
\address{Department of Applied Mathematics\\ University of Twente\\
P.O. Box 217, 7500 AE Enschede, The Netherlands}

\begin{abstract}
Constructive gravity allows to calculate the Lagrangian for gravity, provided one previously prescribes the Lagrangian for all matter fields on a spacetime geometry of choice.
We explain the physical and mathematical foundation of this result and point out how to answer questions about gravity that could not be  meaningfully asked previously.
\end{abstract}


\maketitle

\section{Introduction}
Constructive gravity reveals a connection between matter dynamics and gravitational dynamics, which is deeper than previously appreciated. The key result    \cite{Schuller:2016onj} is this: Three physically mild conditions on a given action 
\begin{equation}
    S_\textrm{\tiny matter}[A_, G)
\end{equation}
for a matter field $A$ (on which the action depends locally) on a geometric background described by one or several tensor fields $G$ (on which the action depends only ultralocally), suffice 
to derive a universal system of linear homogeneous partial differential equations whose coefficient functions depend on the given matter action and whose solutions provide diffeomorphism-invariant actions
\begin{equation}
    S_\textrm{\tiny geometry}[G]
\end{equation}
for the geometry that are causally consistent with the initially prescribed matter dynamics and must be be added to the latter in order to provide a closed system of equations for both the matter and the geometry. 

The mechanism behind this causally consistent closure of given matter field dynamics is simple. First, the matter dynamics unequivocally determine    \cite{Raetzel:2010je} --- no matter how complicated the background geometry on which they are formulated and as long as elementary well-posedness and energy conditions hold --- all ways in which the spacetime may be foliated into initial data hypersurfaces for the matter degrees of freedom. The only further step then consists in a straightforward technical exploitation   \cite{Kuchar:1974es} of the requirement   \cite{Hojman:1976vp} that the geometric background be furnished with canonical dynamics that evolve the pertinent {\it geometric} degrees of freedom between any two of the initial data hypersurfaces for the stipulated {\it matter} dynamics, thus making these the common initial data hypersurfaces for both the geometry and the matter fields. Without this requirement, no sustained prediction for the total matter-geometry dynamics could possibly be made. With this requirement, one has a condition so strong as to typically determine the action for the geometry up to only a few constants of integration. 
After adding the thus obtained action for the geometry to the matter action, the resulting coupled field equations describe how the geometry is influenced by the presence of dynamical matter fields (which amounts to the generation of a gravitational field) and, conversely, how the motion of matter is influenced by the geometry (which amounts to the influence exerted by gravity on matter).   

The thus constituted constructive gravity program hence simply takes a matter action as an input and provides a canonically compatible gravitational action as its output. Starting the program with standard model matter    \cite{WierzbaMSc}, one obtains the Einstein-Hilbert action with two undetermined constants in place of the gravitational and the cosmological constant. Starting the program, instead, with modifications of standard model matter, a correspondingly modified action for the underlying geometry can be obtained and provides the gravity theory for all geometric degrees of freedom that is selected by causal consistency. In any case, gravitational field dynamics are revealed to be a mere consistency condition, imposed by the entirety of matter field dynamics one postulates and tailor-made for the geometric background fields employed by the matter field equations. In programmatic brevity, the philosophy underlying constructive gravity is: Matter first, gravity second. 

\subsection{A previously not solvable problem}
In order to appreciate the far-reaching implications of what is said above, consider the following phenomenologically interesting observational scenario. Assume that future advances in radioastronomy reveal that electromagnetic waves propagating through vacuum regions of space suffer birefringence effects of various strengths, however small, while there is no indication at all that there would be any violation of the linear superposition principle in the observed energy range. This scenario presents one small step for the matter phenomenologist, but one giant leap for the gravitational theorist. It is easy for the matter phenomenologist, since the most general electrodynamics action that generates linear field equations on a tensorial background geometry takes the form   \cite{Rivera:2011rx}
\begin{equation}\label{eq_GLED}
    S_\textrm{\tiny{matter}}[A,G) = \int d^4x\, \omega_G\, G^{abcd} F_{ab} F_{cd}\,,
\end{equation}
where $A$ is the familiar electromagnetic gauge potential with the associated field strength $F=dA$, while $G$ is an, at first arbitrary, fourth-rank tensor field and $\omega_G= (\epsilon_{ijkl} G^{ijkl})^{-1}$ is a weight-one scalar density constructed from it. The quadratic appearance of the field strength $F$ and its two-form character render only those components of $G$ relevant that conform to the algebraic symmetry conditions
\begin{equation}
    G^{abcd} = G^{cdab} \quad \textrm{and} \quad G^{abcd} = - G^{bacd}\,,
\end{equation}
while finiteness of the density factor imposes the open condition $\epsilon_{ijkl} G^{ijkl} \neq 0$. Any observed birefringence effect can now be modelled in one way or another by suitably adapting the 21 independent components of the tensor field $G$ at each point within the spacetime region of the electromagnetic wave such as to fit the obervational data. Given the described hypothetical observations, this is certainly the correct classical matter model    \cite{Hehlbook}. But in this form, it is yet of little predictive power, since both the location and strength of vacuum birefringence and further effects beyond Maxwellian electrodynamics entirely depend on the values taken by the fourth-rank tensor field $G$. Without a way to predict the values taken by the tensor field $G$, one is thus not able to predict the electromagnetic field either. 

The only way to predict the values of the geometric field $G$ (up to equivalence under diffeomorphisms) with the least possible prejudice is to furnish $G$ with dynamics of its own, by extending the action (\ref{eq_GLED}) to the total action
\begin{equation}
    S[A,G] = S_\textrm{\tiny{matter}}[A,G) + S_\textrm{\tiny{geometry}}[G]
\end{equation}
such that the stationarity conditions
\begin{equation}\label{refinedtotaleom}
    \frac{\delta S_\textrm{\tiny{matter}}}{\delta A_a}[A,G) = 0 \qquad \textrm{and} \qquad  \frac{\delta S_\textrm{\tiny{geometry}}}{\delta G^{abcd}}[G] = - \frac{S_\textrm{\tiny{matter}}}{\delta G^{abcd}}[A,G) 
\end{equation}
recover both the phenomenologically enforced general linear electromagnetic field equations and the gravitational field equations. The central problem to solve in order to make this work, of course, is the identification of all physically consistent choices of the action $S_\textrm{\tiny{geometry}}[G]$ for the geometry. But finding the gravitational actions that can underpin given matter dynamics is the very problem solved by constructive gravity. 

\subsection{A familiar problem solved a century ago}
It will be enlightening to see how standard general relativity arises, in the philosophy of constructive gravity, from standard model matter. Other than for the previous example, the reader will not need to have mastered the general machinery described in this article in order to follow the steps of the constructive gravity program in some more detail for the present case, because one simply recovers known concepts from general relativity -- even if from a slightly different conceptual perspective: all mathematical objects of the general theory reduce here to their familiar form.  Simplifying as much as possible for the purpose of clarity, we consider the matter field action 
\begin{equation}\label{KGaction}
    S_\textrm{\tiny matter}[\varphi,g) = \int d^4x \sqrt{\det g} \left(g^{ab} \partial_a \varphi \partial_b \varphi - m^2 \varphi^2\right)
\end{equation}
for a scalar field $\varphi$ and a second rank tensor field $g$, about which we do not need to assume anything a priori beyond the symmetry and non-degeneracy conditions 
\begin{equation}
    g^{ab} = g^{ba} \qquad \textrm{and} \qquad \det g \neq 0\,.
\end{equation} 
In order to find the elementary well-posedness and energy conditions on the matter theory that need to be satisfied for the constructive gravity program to apply, one first calculates the principal polynomial of the postulated matter field equations in each spacetime cotangent space (see section \ref{sec:principal} for an outline of the general theory), which information is equivalent to the one held in an even-rank totally symmetric contravariant tensor field $P$. For the dynamics defined by (\ref{KGaction}), this principal tensor field happens to be of second rank and turns out to be given by
\begin{equation}\label{Pgcoincidence}
    P^{ab} = g^{ab}\,.
\end{equation}
From the point of view of the general theory, it is a pure coincidence that this principal tensor field $P$ has the same rank as the fundamental geometry $g$ and additionally that it is in fact identical to it. (That this truly is a coincidence is impressively illustrated by the fact that for the previously considered general linear electrodynamics, the principal tensor field 
$P^{abcd} = \omega_G^2 \epsilon_{mnpq} \epsilon_{rstu} G^{mnr(a} G^{b|ps|c} G^{d)qtu}$
is found to be cubic in the fundamental geometry $G$ underlying that matter theory and as a totally symmetric tensor certainly does not share its index symmetries, although it coincidentally has the same rank.)  Indeed, some occurrences of the inverse metric in general relativity are in fact occurrences of the principal tensor field, while others are occurrences of the fundamental geometry. Failure to recognize this degeneracy as coincidental goes with impunity only in general relativity proper, but in fact lies at the heart of causality problems of various generalized theories of gravity and matter   \cite{Velo:1970ur}. The general well-posedness and energy conditions   \cite{Schuller:2016onj,Raetzel:2010je} that a matter theory must satisfy in order to be a viable starting point for the constructive gravity program boil down, in the present special example, to the condition that the principal polynomial $P$ be of Lorentzian signature. That also the {\it fundamental geometry} $g$ must have Lorentzian signature is only due to the very particular coincidence (\ref{Pgcoincidence}). 

It is obvious that for the previous subsection's general linear electrodynamics with its pertinent fourth-rank principal tensor $P$, a more general algebraic classification than the signature classification of symmetric bilinear forms needs to kick in and that the implications for the underlying fundamental geometry $G$ are even more intricate to extract, but ultimately obtainable    \cite{Schuller:2009hn}. From a practical point of view, here and in general, one may perfectly ignore the rather extensive theoretical machinery running in the background and write down the the so-called gravitational closure equations (in either functional differential or partial differential form) --- which determine the desired gravitational Lagrangian within the constructive gravity program and are informed by the assumed matter theory --- by ultimately employing only the pertinent fundamental geometry and the calculated principal polynomial in the coefficient functions of the closure equations. Their solution   \cite{Kuchar:1974es,Schuller:2016onj} for the present case of Klein-Gordon dynamics on a Lorentzian background yields -- without any a priori knowledge of metric geometry whatsoever -- the
two-parameter family 
\begin{equation}
    S^{\kappa,\Lambda}_\textrm{\tiny geometry}[g] = \kappa \int d^4x \sqrt{\det g}\left(R - 2 \Lambda\right)
\end{equation}
of gravitational actions, which one recognizes as the Einstein-Hilbert action with both the gravitational and cosmological constant left to be determined by experiment. The Ricci scalar, or rather its definition in terms of the metric, arises automatically in the solution of the closure equations, which are just informed about the matter dynamics though the principal polynomial of the latter and indeed the underlying fundamental geometry. In the parlance of the constructive gravity program, Einstein-Hilbert gravity arises as the gravitational closure of Klein-Gordon theory. The same result is obtained by starting from Maxwell theory or indeed the entire standard model    \cite{WierzbaMSc}. 

\section{Principal polynomials of matter field equations}\label{sec:principal}
\noindent We now turn to an exposition of the general theory, for which the previously mentioned examples present just special cases.

Starting point of the gravitational closure mechanism is a matter action of the form
\begin{equation}
  S_\textrm{\tiny matter}[A, G) = \int_M d^4x \, \mathcal{L}(A(x), \partial A(x), \dots, \partial^N\!\!A(x), G(x))\,,
\end{equation}
where $\mathcal{L}$ is a scalar density, $A$ is a smooth $GL(4,\mathbb{R})$-irreducible tensor field (or, more generally, a finite collection of various such) representing the matter whose dynamics is determined by $S_\textrm{\tiny matter}$ and $G$ is a smooth tensor field  (or, again, a finite collection of various such), to which we will refer as the geometry on the smooth four-dimensional manifold $M$. Note that we assume that our matter actions depend locally on the matter fields and ultralocally on the geometry. 

Variation with respect to the matter field $A$ yields the tensor-density equations of motion
\begin{equation}\label{eom}
   0 = \frac{\delta S_\textrm{\tiny matter}}{\delta A^\mathcal{M}(x)} \equiv \sum_{n=0}^N Q_{\mathcal{MN}}^{a_1\dots a_n}[A(x),G(x)] \partial_{a_1} \cdots \partial_{a_n} A^\mathcal{N}(x)\,,
\end{equation}
where $A^\mathcal{M}(x)$ indicates components with respect to the $GL(4,\mathbb{R})$ representation space in which $A(x)$ takes its values at any point $x\in M$ and the $Q$ are $N+1$ coefficient functionals with local dependence of $A$ and $G$. For field equations that are linear in their highest order derivative term, the corresponding coefficient functions
\begin{equation}\label{naivehighestorder}
   Q^{a_1 \dots a_N}_\mathcal{MN}(G(x))
\end{equation} 
do not depend on the field $A$ at all, and only ultralocally on the geometry $G$. In order to avoid inessential technical complications, we restrict attention here to such matter models.  

We first discuss the case in which the field equations (\ref{eom}) do not feature any hidden integrability conditions, which could otherwise alter the highest derivative coefficient functions. In this straightforward case, the causal structure of the matter field dynamics is encoded entirely in the functions (\ref{naivehighestorder}).  In particular, well-posedness for the matter field dynamics described by $S_\textrm{\tiny matter}$ on a fixed geometry $(M,G)$ --- in other words, the question of whether one can find a suitable foliation of spacetime into hypersurfaces such that prescription of initial field data on one such hypersurface suffices for the dynamics to predict the data on another, neighbouring, such hypersurface such that the resulting spacetime matter field solves the field equations --- requires that the $a$-solution space of the infinite frequency limit Wentzel-Kramers-Brillouin condition
\begin{equation}
    Q^{a_1 \dots a_N}_\mathcal{MN}(G(x))k_{a_1} \dots k_{a_N} a^\mathcal{N} = 0
\end{equation}
is at least $(S+1)$-dimensional, where $S$ is the dimension of the gauge orbits featured by the matter theory     \cite{Schuller:2016onj}. It can be shown    \cite{Hehl:2002hr,Itin:2009aa} that this condition can always be written as a polynomial condition
\begin{equation}\label{specialPtilde}
   \widetilde P^{i_1 \dots i_{\deg \widetilde P}}(x) k_{i_1} \cdots k_{i_{\deg \widetilde P}} = 0 \qquad \textrm{for } k \in T_x^*M\,,
 \end{equation}
where $\widetilde P^{i_1 \dots i_{\deg P}}(x)$ are the coefficient functions for some totally symmetric tensor density $P$ of weight one. Since this is a homogeneous condition we may and will regularly de-densitize this condition by use of some meaningful scalar density that can be constructed ultralocally in terms of the geometry $G$ and then denote the resulting totally symmetric tensor field by $P$. Since, in the language of the theory of partial differential equations, $P(x)$ is the principal polynomial (in the cotangent fibre variable $k$) of the field equations (\ref{eom}), we refer to $P$ as the principal tensor field.

In the presence of hidden integrability conditions -- which are only revealed by systematically repeated differentiation and elimination of the equations that were originally obtained by variation -- the highest order coefficient functions may be altered once the integrability conditions have been made explicit. The simplest example illustrating this is the system 
\begin{equation}\label{samplenoninvolutivesystem}
A_x + A_{yy} = 0\qquad \textrm{and} \qquad A_y + A_{yx} = 0\,,
\end{equation}
for which only differentiation and elimination reveals the contained implication $A_{xx} - A_{yy} = 0$, which crucially alters the highest order coefficient functions of the equations of motion  (even making them into a non-square matrix) and hence the calculation and final form of the principal tensor. 

Extending a given system of partial differential equation, such that all hidden integrability conditions are made explicit, is achieved by the Cartan-Kuranishi algorithm    \cite{Kuranishi:1951}. From the highest order coefficient functions of a so obtained system, which is then termed involutive, one may then calculate the principal tensor, by a slight generalization of the prescription given before for systems without hidden integrability conditions. The algorithm for taking equations of motion $\Phi_B[A] = 0$ for fields $u^\mathcal{M}$ into involutive form revolves around repeated calculation of the so-called geometric symbols
\begin{equation}
  (M_q)_{B\mathcal{N}}{}^\nu := \frac{\partial \Phi_B}{\partial A^\mathcal{N}{}_{\!,\nu}}[A]\qquad\textrm{ for } q = \nu_1+\dots+\nu_{\dim M}\,,
\end{equation}
where the derivative $A^\mathcal{N}{}_{\!,\nu}$ with respect to the multi-index $\nu = (\nu_1, \dots, \nu_{\dim M })$ denotes the $q$-th partial derivative $\partial_1^{\nu_1} \partial_2^{\nu_2} \dots \partial_{\dim M}^{\nu_{\dim M}} A^\mathcal{N}$ of the field component $A^\mathcal{N}$, for $q$ being the currently highest derivative order of the intermediate system of equations generated in each step of the now easily performed Cartan-Kuranishi algorithm:
\begin{enumerate}
  \item Having calculated the components of the the geometric symbol for the current set of equations (starting with the initially given set of equations if no other set has been generated yet by the algorithm), they are arranged into a matrix $M_q$ whose rows are labeled by the index $B$ and whose columns are labelled by the combination of the indices $\mathcal{N}$ and $\nu$. The only rule for how this labelling is done is that the resulting column indices (non-strictly) decrease in the class $1\leq c(\nu) \leq \dim M$ of the multiindex $\nu$, which is defined as the smallest $i$ for which $\nu_i$ is non-zero. The actual calculational step then consists in taking the thus constructed matrix to row echelon form by judicious linear combinations of row vectors only. For any $i=1,\dots,M$ one then reads off the set of coefficients 
\begin{equation}
    \beta_q^{(i)} := \textrm{number of pivot elements in all columns of class $i$}\,. 
\end{equation}  
\item Prolongate the current system, i.e., combine it with all $\dim M$ possible first order partial derivatives of each of its current equations and calculate the matrix
$M_{q+1}$ for this prolongated system. This allows to determine whether the system, as it was before this last prolongation, is pre-involutive. This is the case if the beta coefficients satisfy the pre-involutivity condition
\begin{equation}
\sum_{i=1}^{\dim M} i \beta_q^{(i)} = \textrm{rank}(M_{q+1})\,.
\end{equation}
If this is not the case, consider the just calculated prolongated system the new current system and return to the first step with the thus updated system of equations. It is guaranteed that the above equality will be satisfied after a finite number of iterations on steps 1 and 2, in which case one then proceeds to step 3.
\item Consider the prolongated system that has just been calculated to confirm that the pre-involutivity condition of step 2 has been satisfied, but still consider the unprolongated system as the current system. If no integrability condition (an equation of equal or lower derivative order than the current system that is linearly independent of the latter) can be derived from the prolongated systems, the current system is called involutive and the algorithm terminates. If, however, integrability conditions are found, they are appended to the current system and the such extended system is handed as the new current system to step 1 of the algorithm. It is guaranteed that after a finite number of iterations of steps 1, 2 and 3, the algorithm terminates.   
\end{enumerate}
Application of the algorithm to the system (\ref{samplenoninvolutivesystem}) yields beta-coefficients $\beta_2^{(1)} = 2$ and $\beta_2^{(2)} = 1$ and rank $3$ for the prolongated system, which identifies the initially given system as already pre-involutive in step 2. Since the prolongated system, however, turns up the integrability conditon $A_{xx} - A_{yy} = 0$, the system is not involutive yet. Adding the integrability condition to the original system and repeating steps 1 and 2 one obtains the new beta coefficients $\beta'{}_2^{(1)} = 2$ and $\beta'{}_2^{(2)} = 1$ and rank $4$ for the prolonged system. Thus the original system extended by the found integrability condition is found pre-involutive in step 2 and one indeed finds no further integrability condition in step 3. Thus
\begin{equation}\label{sampleinvolutive}
A_x + A_{yy} = 0\qquad \textrm{and} \qquad A_y + A_{yx} = 0 \qquad {and}\qquad A_{xx} - A_{yy} = 0 
\end{equation}
is the involutive form of the original system (\ref{samplenoninvolutivesystem}) as obtained by the Cartan-Kuranishi algorithm.

For equations of motion $\Phi_B[A]=0$ that follow from a matter action by variation, but are not already involutive, a slight adaptation of the calculation of the principal tensor is required, since the principal symbol 
\begin{equation}
   T_{\overline{B}\mathcal{N}}(k) := \sum_{\nu_1 + \dots + \nu_{\dim M} = \overline{q}}\frac{\partial \Phi_{\overline{B}}}{\partial A^\mathcal{N}{}_{\!,\nu}} (k_1)^{\nu_1} \cdots (k_{\dim M})^{\nu_{\dim M}}\,
\end{equation}
of their involutive form $\Phi_{\overline{B}}[A]=0$ ---
where the index $\overline{B}$ now not only covers the range of the original $B$ but also all the additional equations that had to be added in order to achieve involutive form and $\overline{q}$ is the highest derivative order encountered in the involutive system ---  generically constitutes a non-square matrix $T(k)$. In any case, the principal tensor density can be shown the be determined in this case by taking the determinant of the Gramian matrix of $T(k)$, 
\begin{equation}\label{generalPtilde}
  \widetilde P^{a_1 \dots a_{\deg \widetilde P}} k_{a_1} \cdots k_{a_{\deg \widetilde P}} := \det(T^t(k) T(k))\,.
\end{equation}
Note that the Gramian matrix is a square matrix whose rows and columns are labeled by some $GL(4,\mathbb{R})$ representation, so that its determinant transforms as a scalar density of the appropriate weight. 

For technical reasons, and since it does not affect the information encoded in the principal scalar density $\widetilde P(k)$, whether obtained from (\ref{specialPtilde}) or more generally from (\ref{generalPtilde}), we will not only routinely de-densitize it by multiplication with a suitable density factor, but also reduce its degree as much as possible by dropping repeated factors, so that 
\begin{equation}
  \widetilde P(k) = \omega_G^m P_1^{n_1}(k) P^{n_2}_2(k) \cdots P_F^{n_F}(k)\,,
\end{equation}
where $m$ is an integer and $n_1, \dots, n_F$ are positive integers while $P_1(k), \dots, P_F(k)$ are irreducible polynomials in $k$ that transform as scalar fields for any substitution of a covector field $k$, gives rise to the reduced principal tensor $P$ uniquely determined by 
\begin{equation}\label{irred}
   P^{a_1 \dots a_{\deg P}} k_{a_1} \dots k_{a_{\deg P}} := P_1(k) P_2(k) \cdots P_F(k)\,.
\end{equation}
Only this reduced form of the principal tensor $P$ will play a role in constructiuve gravity and will simply be called the principal tensor of the matter field equations in the following.

\section{Kinematics induced by matter field equations}
\noindent Three technical conditions on any given matter field equations must be satisfied in order to derive a complete kinematical interpretation for the underlying spacetime geometry. They all concern the principal tensor $P$ of the matter field equations obtained from an action $S_\textrm{\tiny matter}[A,G)$, and thus implicitly impose conditions on the geometric tensor $G$ in terms of which $P$ is expressed. Physically, these conditions correspond to classically hardly negotiable necessary requirements for the field equations and their geometric optical limit, namely that (a) there exist initial data surfaces for the field equations, (b) the spacetime structure can be interpreted in a temporal-spatial way and (c) that the spacetime structure allows for time-orientability and corresponding energy-orientability. Technically, these physical conditions amount to
\begin{enumerate}
  \item[(a)] The homogeneous polynomial $P_x: T^*_xM \to \mathbb{R}$ defined in terms of the principal tensor by $k \mapsto P^{a_1 \dots a_{\deg P}}(x) k_{a_1} \cdots k_{a_{\deg P}}$ at each point $x$ of the manifold $M$ must be hyperbolic. This means, by definition, that there exists some $h\in T_x^*M$ with $P_x(h)\neq 0$ such that for all $q \in T_x^*M$ the equation
 \begin{equation}\label{hyperbolicity}
   P_x(q + \lambda h) = 0
 \end{equation}
 possesses $\deg P$  real solutions $\lambda_1, \dots, \lambda_{\deg P}$, counting algebraic rather than geometric multiplicity. One can show that if $h$ satisfies the above condition, then there is an entire connected set of such vectors which constitute an open and convex cone $C_x(h)$, the so-called hyperbolicity cone of $P_x$ that contains $h$. Note that the hyperbolicity cones are uniquely determined by the principal tensor and that the latter has been uniquely constructed in terms of the spacetime geometry $G$, in a way that is crucially informed by the particular equations of motion of the matter field $A$, but is functionally independent of the latter. 
 
Clearly, one can choose any other covector $h'$ in this hyperbolicity cone as an alternative representative, since $C_x(h) = C_x(h')$. Moreover, since $-h$ satisfies condition (\ref{hyperbolicity}) if and only if $h$ does, but since $P(h)\neq 0$, $-h$ does not lie in the same connected component as $h$ and thus $C_x(h) \cap C_x(-h) = \emptyset$; thus there is always an even number of hyperbolicity cones at each point of spacetime. A time-orientation of the spacetime is chosen  by prescription of some smooth and everywhere hyperbolic covector field $n$, which singles out one particular cone $C_x := C_x(n_x)$ at spacetime point $x$. We call these cones the local observer cones, since they contain all possible conormals (which set one may geometrically think of as all possible tangent hyperplanes) to initial data hypersurfaces through the respective spacetime point to which they are attached. This, indeed, is the relation to the question of well-posedness of the field equations. 

We impose a positive sign convention, which will come in handy later on, namely that $P_x(C_x) > 0$ for all $x\in M$. If this does not hold in the first place, then we necessarily have  $P_x(C_x)<0$ for all $x\in M$, due to the continuity of the time-orienting hyperbolic covector field $n_x$ and the continuity of the principal tensor field $P$, so that replacing the principal tensor $P$ by $-P$, which has no effect on the hyperbolicity condition, arranges for the desired sign. 
    
  \item[(b)] The dual polynomial $P_x^\#: T_xM \to \mathbb{R}$, which is uniquely determined by a given hyperbolic polynomial $P_x: T_x^*M \to \mathbb{R}$ up to a non-zero factor (which turns out to cancel for any use the dual polynomial is put to) must be hyperbolic. The dual polynomial $P_x^\#$ of a hyperbolic polynomial $P_x=(P_1)_x \cdots (P_F)_x$, which is decomposed into mutually different irreducible factor polynomials $P_1,\dots, P_F$, is defined as the product 
\begin{equation}
   P_x^\# := (P_1)^\#_x \cdots (P_F)^\#_x\,,
\end{equation}
of the duals of these irreducible factors. Thus it suffices to define the dual polynomial $Q^\#$ associated with an irreducible polynomial $Q$, namely as the likewise irreducible polynomial for which
\begin{equation}\label{dualdefining}
  Q_x^\#(DQ_x(k)) = 0 \quad \textrm{ holds for all } k \in T_x^*M \textrm{ with } Q_x(k)=0 \textrm{ and } DQ_x(k) \neq 0\,,
\end{equation}
where $DQ_x$ denotes the derivative of $Q_x$ with respect to the cotangent space fibre at $x$. The very existence of the dual polynomial $P_x^\#$ hinges on the hyperbolicity of $P_x$, which is equivalent to the hyperbolicity of each irreducible factor polynomial. The immediate physical relevance of the dual polynomial is revealed by the real projective relation 
\begin{equation}
   [DP_x^\#([DP_x([k])])] = [k] 
\end{equation}
for all $k \in T_x^*M$ with $P_x(k)=0$ and $DP_x(k) \neq 0$ and $DDP_x(k)\neq 0$,
where $[\cdot]$ denotes projective equivalence, since it reveals that any projective $P_x$-null covector $[k]$ (safe such on an exceptional subset of measure zero) is bijectively mapped to a projective vector $[DP_x([k])]$, with the inverse map given by $[DP_x^\#]$. Due to the generic non-linearity of these maps, this is highly non-trivial. In physics language, however, it establishes a easily understood fact: for each $P^\#$-null ray direction, which emerges in the geometric-optical limit of the underlying hyperbolic matter field theory, there is a  
unique $P_x$-null wave surface and vice versa. We will refer to $P_x$-null covectors also as massless momenta.


\item[(c)] The positive energy cone $E_x^+$ at each point $x$ of the spacetime $M$ is defined for any hyperbolic principal polynomial $P_x$ with hyperbolic dual polynomial $P_x^\#$, as required by conditions (a) and (b), as the closed convex cone 
\begin{equation}
   E_x^+ := \{ e \in T_x^*M \,|\, e(X) > 0 \textrm{ for all } X \in \ell_x^{-1}(C_x)\}\,. 
\end{equation}
The physical rationale for this definition is clear: a massless or massive momentum $p$ can only be said to be of positive energy if all observers (represented here by all possible oberserver worldline tangent vectors $X$)   agree on the sign of the respectively seen energy $e(X)$. 
Any massive momentum $p$, as defined under (b) above, is automatically of positive energy, by construction of $C_x^\#$. 

The final condition on the principal polynomial, and thus the underlying spacetime geometry $G$, is that any massless momentum $p$ at some point $x\in M$ must lie either in the positive energy cone $E_x^+$ or the negative energy cone $E_x^-$. This energy distinguishing condition is physically required, since it is necesary to have all observer agree on whether a decay that involves a massless particle is kinematically possibly or not. 

Together with the previously adotped hyperbolicity condition on the dual polynomial $P_x^\#$, the energy distinguishing conditions allows to unquely identify the hyperbolicity cone $C_x^\# \subset T_xM$ of $P_x^\#$ for which the observer cone $C_x \subseteq \ell_x(C_x^\#)$, where $\ell_x$ denotes the invertible Legendre map
\begin{equation}
   \ell_x: C_x^\# \to \ell_x(C_x^\#) \subset T_x^*M\,,\qquad X \mapsto - \frac{1}{\deg P_x^\#} \frac{DP_x^\#(X)}{P_x^\#(X)}\,.
\end{equation} 
The corresponding restriction of the inverse $\ell_x^{-1}$ of this map to the observer cone $C_x$ is physically easily understood as the bijective map between the massive momenta $p$ of mass $m$ at the point $x\in M$, which are characterized by $p \in C_x$ and $P_x(p)=m^{\deg P}$ for some positive mass  $m$, and the tangent vectors $\ell_x^{-1}(p)$ of their respective worldlines. 
\end{enumerate}
Matter dynamics satifying the above three conditions impose the kinematical interpretation of the spacetime geometry $(M,G)$, with the relevant information coming from the matter dynamics being encoded in the principal polynomial. The three kinematical constructions of immediate practical importance are
\begin{enumerate}
\item A local observer is given by a curve $e: (a,b) \longrightarrow LM$ in the spacetime frame bundle $\pi: LM \stackrel{\pi}{\longrightarrow} M$ such that {\it (i)} the first frame vector $e_0(\lambda)$ lies in the Legendre dual $\ell_{(\pi\circ e)(\lambda)}^{-1}(C_{(\pi\circ e)(\lambda))})$ of the cotangent space observer cone for all $\lambda \in (a,b)$ and  {\it (ii)} the other frame vectors $e_1(\lambda), e_2(\lambda), e_3(\lambda)$ are Legendre-orthogonal to $e_0(\lambda)$, which is to say that they lie in the kernel of
$\ell_{(\pi\circ e)(\lambda)}(e_0(\lambda))$ for every $\lambda \in (a,b)$. 
Physically, this means that for every point of every initial data hypersurface, one can find a local observer whose worldline $\pi \circ e$ pierces the hypersurface at this point and whose worldline tangent vector $e_0$ is Legendre orthogonal to the hypersurface's tangent directions. These latter tangent directions are the purely spatial directions seen by this particular observer. 

\item A first order action for the worldline $x: \mathbb{R} \longrightarrow M$ of a massless point particle is immediately implied by the dispersion relation $P(k)=0$ that must hold for the momentum of such a particle, namely
\begin{equation}
   S_\textrm{\tiny massless}[x,k,\mu] = \int d\lambda \,[k_a(\lambda) \dot x^a(\lambda) - \mu(\lambda) P_{x(\lambda)}(k(\lambda))]\,,
\end{equation}
where $\mu$ is a Lagrange multiplier. But solving the corresponding equations of motion requires solving for $k$. By virtue of the inverse Gauss map $[DP^\#]$ and the homogeneity of $P^\#$, one can indeed solve for $k = \sigma DP^\#$ in terms of another Lagrange multiplier $\sigma$ (which absorbs the projective scaling ambiguity). Encoding this elimination directly into the action one finds the equivalent second oder action
\begin{equation}
  S_\textrm{\tiny massless}[x,\sigma] = \int d\lambda\, \sigma(\lambda) P_{x(\lambda)}^\#(\dot x(\lambda))\,,
\end{equation}
which can be straighforwardly varied without knowledge of the generically non-linear kinematical machinery running in the background.

\item For massive point particles, the same philosophy applies, but entirely different mathematics are at work. Instead of projective algebraic geometry and projective Gauss maps, as for the massless particle, it is now convex analysis and the Legendre map and its inverse that play the crucial role. Also in this case,
an obvious first order action
\begin{equation}
  S_\textrm{\tiny massive}[x,k,\mu] = \int d\lambda \,\left[k_a(\lambda) \dot x^a(\lambda) - \mu(\lambda) (\ln P_{x(\lambda)})\left(\frac{k(\lambda)}{m}\right)\right]
\end{equation}
leads to the problem of inverting a non-linear velocity-momentum relation, which is now achieved by virtue of the Legendre map $k = m \ell_x(\dot x/(\lambda \deg P^\#))$, which can be used to arrive at the equivalent second order action
\begin{equation}
  S_\textrm{\tiny massive}[x] = \int d\lambda \,P(\ell_{x(\lambda)}(\dot x(\lambda)))^{-1/(\deg P^\#)}\,.
\end{equation}
\end{enumerate}
It is instructive to note how the kinematics of standard general relativity follow from the above general theory, starting from even the simplest possible field dynamics one wishes to have available on a Lorentzian metric manifold of signature $(+---)$,
\begin{equation}
     S_\textrm{\tiny KG matter}[\phi] = \int d^4x \sqrt{-\det g(x)} \left(g^{ab}(x) \partial_a \phi(x)\partial_b\phi(x) - m^2 \phi^2(x)\right)\,
\end{equation} 
whose equations of motion yield a second rank principal tensor field with components $P^{ab} = g^{ab}$, so that hyperbolicity of $P$ is equivalent to the supposed Lorentzian signature of the metric. The hyperbolicity cones of $P_x$ at  each point  $x$ are two disjoint open convex cones of covectors $k$ for which $g_x^{ab}>0$ and a time-orientation identifies one of them as the observer cone $C_x$. The positive sign convention $P_x(C_x)>0$ is immediately satisfied because of the mainly minus signature chosen for the metric. The dual polynomial is $P^\#_{ab} = g_{ab}$ and the projective Gauss maps $[D^aP(k)] = [g^{ab} k_b]$ and $[D_a^P\#(X)] = [g_{ab} X^b]$ simply raise and lower the index and are linear in this simple case. The positive energy cone is the closure $\overline{C_x}$ of the observer cone, and thus indeed captures all massive momenta contained in $C_x$ and all massless momenta on the boundary $\partial C_x$ that is not part of the open cone $C_x$ and none of the momenta in $-\overline{C_x}$ are captured. The Legendre map and its inverse evaluate to
$(\ell_x(X))_a = - g_{ab} X^b / g_{mn} X^m X^n$ and $(\ell^{-1}_x(k))^a = - g^{ab} k_b / g^{mn} k_m k_n$, yielding precisely the velocity-momentum relation for massive particles and reveal the generically required Legendre orthogonality of purely spatial observer frame vectors to the temporal frame vector as simply their Lorentzian metric orthogonality in this simple case. The general action for massless and massive particles reduces to the known actions in general relativity,
$$S_\textrm{\tiny massless}[x,\sigma] = \int d\lambda\, \sigma(\lambda) g_{ab}(x(\lambda)) \dot x^a(\lambda) \dot x^b(\lambda)$$
and
$$S_\textrm{\tiny massive}[x] = \int d\lambda \,\sqrt{g_{ab} x^a(\lambda) \dot x^b(\lambda)} \,.$$

The general theory presented before generalizes these cornerstones of general relativity to generalized tensorial spacetimes, in a way that is informed by the specific matter dynamics one stipulates on the given spacetime. From this point of view --- which is indeed the view taken by Einstein when he distilled the kinematical lessons conveyed by Maxwell dynamics into the spacetime structure and its interpretation --- the kinematical interpretation of a spacetime geometry $(M,G)$ cannot be extracted from, or be assigned to, the geometry per se. Indeed, had we not chosen Klein-Gordon theory (or any other standard model field) as the matter field theory on our Lorentzian manifold, but instead, say, a Proca theory with quartic self-interaction, we would have obtained a vastly different kinematical interpretation of the very same Lorentzian manifold.
In essence, the kinematics impressed on a geometry follows only from the triple $(M,G,S_\textrm{\tiny matter}[A,G))$. This is an insight that was so far essentially ignored in the overwhelming majority of attempts to construct gravity theories beyond general relativity. 

\section{Gravitational closure}\label{sec_gravclosure}
Consider a foliation of the spacetime $M$ into initial data surfaces, described by a one-real-parameter family of smooth embedding maps $X_t: \Sigma \longrightarrow M$, where $\Sigma$ is a smooth three-dimensional manifold such that  the image $X_t(\Sigma)$ is an initial data surface for the matter theory. We now define projection frames by choosing coordinate maps $y^\alpha$, for $\alpha = 1,\dots, 3$ on $\Sigma$, whence we obtain a spacetime tangent basis 
\begin{equation}\
    e_{0}(t,\sigma) = \ell_{X_t(\sigma)}^{-1}(n_t(\sigma)) \qquad \textrm{and}\qquad e_{\alpha}(t,\sigma) = X_{t*}(\left(\frac{\partial}{\partial y^\alpha}\right)_{\!\!\sigma})
\end{equation}
along each embedded hypersurface $X_t(\Sigma)$, where each $n(t,\sigma)$ is a spacetime covector field normal to the hypersurface and normalized with respect to $P^\#$,
\begin{equation}
   n(t,\sigma)(e_{\alpha}(t,\sigma)) = 0 \qquad\textrm{and}\qquad P_{X_t(\sigma)}^\#(\ell_{X_t(\sigma)}^{-1}(n(t,\sigma)) = 1\,.
\end{equation}
Employing this frame $e_a(t,\sigma)$ and the unique dual frame $\epsilon(t,\sigma)$, we obtain an obvious projection of the spacetime geometry $G$ to several one-real-parameter families of induced tensor fields on $\Sigma$. For a geometry dexcribed by a $(1,3)$-tensor field,  for instance, we obtain eight one-real-parameter families of tensor fields on $\Sigma$ with valence $(0,0)$, $(0,1)$, $(0,1)$, $(0,2)$, $(1,0)$, $(1,1)$, $(1,1)$ and $(1,2)$, namely
\begin{equation}
  \mathbf{g}_t^{{}^p{}_{qr}}(\sigma) = G_{X_{t}(\sigma)}(\epsilon^p(t,\sigma), e_q(t,\sigma),e_r(t,\sigma))\,
\end{equation}
for ${}^p{}_{qr} \in \{ {}^0{}_{00}, {}^0{}_{0\gamma}, {}^0{}_{\beta 0}, {}^0{}_{\beta\gamma}, {}^\alpha{}_{00}, {}^\alpha{}_{0\gamma}, {}^\alpha{}_{\beta 0}, {}^\alpha{}_{\beta\gamma}  \}$.
 Analogously for geometric tensor fields of different valence or even several such tensor fields of various valences. It proves useful to notationally collect the occurring index combinations in one caligraphy index $\mathcal{A}$, such that we can write $\mathbf{g}_t^\mathcal{A}(\sigma)$.  Analogously, the dual $P^\#$ of the principal tensor field yields $\deg P^\# + 1$ one-parameter families of totally symmetric covariant tensor fields $\mathbf{p}_{t\,\alpha_1 \dots \alpha_n}(\sigma)$ for $n=0, 1, \dots, \deg P^\#$. An important point is that our projection frames are constructed such that the first two projections always take the values
\begin{equation}\label{frameconditions}
   \mathbf{p}_t(\sigma) = 1 \qquad \textrm{and}\qquad \mathbf{p}_t{}^{\alpha}(\sigma) = 0\,. 
\end{equation}
But since the dual of the principal polynomial is given ultralocally in terms of the spacetime geometry, one can write these two induced tensor fields, in particular, as functions of the induced geometric fields,
\begin{equation}
 \mathbf{p}_t(\sigma) = p(\mathbf{g}_t^\mathcal{A}(\sigma)) \qquad \textrm{and} \qquad \mathbf{p}_t{}^\alpha(\sigma) = p^\alpha(\mathbf{g}_t^\mathcal{A}(\sigma))\,.
\end{equation}
Thus the properties (\ref{frameconditions}) impose a (generically non-linear) algebraic relation between the projected fields $\mathbf{g}_t^\mathcal{A}$. While these are automatically satisfied once the definition of the $\mathbf{g}_t^\mathcal{A}$ are employed, they become non-trivial if one turns to the canonical view of dynamics, which no longer considers the spacetime geometry $G$ as the fundamental variables of the theory and the projected fields $\mathbf{g}_t^\mathcal{A}$ as derived quantities, but precisely the other way around. This change in perspective is reflected by introducing tensor fields $g_t^\mathcal{A}$ on $\Sigma$ whose tensorial structure mimics that of the projections $\mathbf{g}_t^\mathcal{A}$, including the algebraic index symmetries the projections once inherited from the spacetime geometry, but are functionally no longer related to the spacetime geometry. But then the generically non-linear conditions (\ref{frameconditions}) must be imposed explicitly as
\begin{equation}
  p(g_t^\mathcal{A}(\sigma)) = 1 \qquad \textrm{and} \qquad p^\alpha(g_t^\mathcal{A}(\sigma)) = 0\,,
\end{equation}
since they no longer follow automatically. But instead of dealing with such non-linear constraints, we introduce generalized configuration fields  $\varphi^1, \dots, \varphi^F$ on $\Sigma$ and parametrization maps $\widehat{g}^\mathcal{A}(\phi^1,\dots,\phi^F)$ such that the tensor fields 
\begin{equation}\label{fieldparam}
  g_t^\mathcal{A} = \widehat{g}^\mathcal{A}(\phi_t^1,\dots,\phi_t^F)
\end{equation}
generated from from one-real-parameter families of these configuratiomn fields
satisfy the linear symmetry conditions and generically non-linear frame conditions (\ref{frameconditions}) while the configuration variables diffeomorphically parametrize the remaining degrees of freedom. The latter is ensured by requiring also the existence of inverse maps $\widehat \varphi^A$ with for $A=1,\dots, F$, such that
\begin{equation}
  \widehat g^\mathcal{A}(\widehat{\varphi}^A(g)) = g^\mathcal{A}
      \qquad\textrm{and} \qquad     \widehat\varphi^A(\widehat{g}^\mathcal{A}(\varphi) = \varphi^A \,,
\end{equation}
from which the important relation
\begin{equation}
   \frac{\partial \widehat\varphi^A}{\partial g^\mathcal{A}}(\widehat g(\varphi)) \frac{\partial \widehat g^\mathcal{A}}{\partial\varphi^B}(\varphi) = \delta^A_B 
\end{equation}
follows. 

With the above preparations made, we can now calculate the two coefficient functions $F^A{}_\mu{}^\gamma(\varphi)$ and $M^{A\gamma}(\varphi)$ of the countable set of linear homogeneous partial differential equations that must be solved in order to obtain the gravitational actions that are causally consistent with the given matter field dynamics. The first of these coefficients can be read off the right hand side of 
\begin{equation}
     (\mathcal{L}_{\vec{N}} \widehat g)^\mathcal{A}(\varphi)  \frac{\partial \widehat\varphi^A}{\partial g^\mathcal{A}}(\widehat g(\varphi)) =: N^\mu \partial_\mu \varphi^A + \partial_\gamma N^\mu F^A{}_\mu{}^\gamma(\varphi)\,,
\end{equation}
where $\vec N$ is some vector field on $\Sigma$. The second one is calculated directly from
\begin{equation}
   M^{A\gamma}(\varphi) =  
   \frac{\partial \mathbf{g}^\mathcal{A}}{\partial\partial_\gamma X^a} e_0^a
   \frac{\partial \widehat\varphi^A}{\partial g^\mathcal{A}}(\widehat g(\varphi))\,,
\end{equation} 
which is easily calculated by expressing $\mathbf{g}^\mathcal{A}$ in terms of the spacetime geometry $G$ (for which $\partial G/\partial\partial_\gamma X^a$ vanishes) and using the relations
\begin{eqnarray}
   \frac{\partial e_0^m}{\partial\partial_\gamma X^a} &=& \frac{1}{1- \deg P^\#} e_\sigma^m p^{\sigma\gamma}(\widehat g(\varphi))\,, \\
    \frac{\partial e_\mu^m}{\partial\partial_\gamma X^a} &=& \delta^m_a \delta^\gamma_\mu\,,\\
    \frac{\partial \epsilon_m^0}{\partial\partial_\gamma X^a} &=& - \epsilon^0_a \epsilon^\gamma_m\,,\\
    \frac{\partial \epsilon_m^\mu}{\partial\partial_\gamma X^a} &=&  - \epsilon^\mu_a \epsilon^\gamma_m + \frac{1}{\deg P^\# - 1} \epsilon^0_m \epsilon^0_a p^{\mu\gamma}(\widehat g(\varphi))\,.
\end{eqnarray}

The key result    \cite{Schuller:2016onj} of constructive gravity is that the gravitational dynamics for the spacetime geometry are given by the action
\begin{equation}\label{gravityaction}
   S_\textrm{\tiny geometry}[G(\varphi_t,N_t,\vec N_t)] = \int d^4x\, N \mathscr{L}\left(\varphi_t, K[\varphi_t,N_t,\vec N_t]\right)
\end{equation}
where 
\begin{equation}
   K^A[\varphi,N,\vec N] := \frac{1}{N} \Big(\dot\varphi - (\partial_\gamma N)  M^{A\gamma}[\varphi] - N^\mu \partial_\mu \varphi^A + (\partial_\gamma N^\mu) F^A{}_\mu{}^\gamma(\varphi) \Big)
\end{equation}
and the scalar density $\mathscr{L}$ of weight one is a functional
of the $\varphi$ and a function of the $K$ that is determined by the gravitational closure equations, which in functional differential form are just the two following equations 
\begin{eqnarray}
	0 &=& - K^B(y)  \frac{\delta \mathscr{L}(x)}{\delta\varphi^B(y)} + \left(\partial_\gamma \delta_x\right)(y) K^B(y) {M^{A\gamma}{}_{:B}}(x) \frac{\partial \mathscr{L}}{\partial K^A}(x) + \partial_\mu \left( \frac{\delta \mathscr{L}(x)}{\delta\varphi^B(\cdot)} {M^{B\mu}{}} \right)(y) \nonumber \\
	& & + \partial_\mu \frac{\partial \mathscr{L}}{\partial K^A}(x) \Big[  (\mathrm{deg} P^\# -1)^{-1} {p^{\rho\mu}}{F^A{}_\rho{}^{\nu}} - {M^{B[\mu|} M^{A|\nu]}{}_{:B}} \Big](x) \left(\partial_\nu\delta_x\right)(y) \nonumber\\
	& & - \frac{\partial \mathscr{L}}{\partial K^A}(x) \Big[ (\mathrm{deg} P^\#-1)^{-1} {p^{\rho\nu}}\left(\partial_\rho \varphi^A + {F^A{}_\rho{}^\gamma{}_{,\gamma}}\right)  + \partial_\mu \left( {M^{B[\mu|}}{M^{A|\nu]}{}_{:B}} \right) \Big] (x) \left(\partial_\nu\delta_x\right)(y) \nonumber\\ 
    & &- (x \longleftrightarrow y). \label{functionaleqn1}
\end{eqnarray}
and
\begin{eqnarray}
	0 &=& \frac{\partial \mathscr{L}}{\partial K^B}(y)\, K^A(y) \left(\delta^B_A \delta^\gamma_\mu  + {F^B{}_{\mu}{}^\gamma{}_{:A}} \right)(y) (\partial_\gamma \delta_y)(x) 
    - K^A(y)  \partial_\gamma \frac{\partial \mathscr{L}}{\partial K^B}(y) {F^B{}_{\mu}{}^\gamma{}_{:A}} (y) \delta_y(x) \nonumber \\
    & & - \left( K^A  \frac{\partial \mathscr{L}}{\partial K^A} - \mathscr{L}\right)(y) (\partial_\mu \delta_y)(x) + \partial_\mu \left( K^A  \frac{\partial \mathcal{L}}{\partial K^A} - \mathscr{L} \right)(y) \delta_y(x) \nonumber \\
    & & + \left(\partial_\mu\varphi^A + {F^A{}_\mu{}^\gamma{}_{,\gamma}}\right) (x) \frac{\delta \mathscr{L}(y)}{\delta \varphi^A(x)} + {F^A{}_\mu{}^\gamma} (x) \partial_\gamma \left( \frac{\delta \mathscr{L}(y)}{\delta \varphi^A(\cdot)} \right)(x)\,,\label{functionaleqn2}
\end{eqnarray}
 where the shorthand $Q_{:A}{}^{\alpha_1\dots \alpha_N} := \partial Q/ \partial\partial_{\alpha_1\dots \alpha_N}\varphi^A$ has been used in both. Solving these equations for $\mathscr{L}$ then completely determines the gravitational action (\ref{gravityaction}) that provides dynamics for the geometry employed in the matter action which is causally consistent with the initially stipulated matter field dynamics. 
 
\section{Solution techniques for gravitational closure equations} 
\noindent\textbf{\textit{General gravitational closure.}}
While it is straightforward to set up the gravitational closure equations for any matter field action on any tensorial background geometry -- as long as the principal polynomial of the resulting matter field equations of motion satisfy the three physicality conditions, which may require restriction of the geometry -- it appears generically prohibitively hard to solve this countable set of linear homogeneous partial differential equations. 

A notable exception is provided by standard model matter on a metric background. The restriction on the geometry, which makes the standard model fields satisfy the physicality conditions, boils down to the metric having Lorentzian signature. In this case, the closure equations are not only set up as swiftly as in any other case, but they can also be solved without further assumptions. 

Maybe the difficulty to solve the gravitational closure equations for other matter models of physical interest, such as general linear -- and thus birefringent -- electrodynamics, is in part due to an unfortunate choice of field parametrization (\ref{fieldparam}) of the field degrees of freedom. A well-considered choice of  parametrization might render a general solution less difficult.

The most promising line of attack, however, is to better understand the structure of the closure equations themselves. A theoretically invaluable step would be to take them to involutive form, maybe in general and maybe case by case. At any rate, this would allow, for instance, a direct calculation of the dimension of their linear solution space. This dimension would then of course be equal to the number of physical constants that remain undetermined by constructive gravity and must hence be measured in experiments. While this number is $2$ in the case of standard model matter -- corresponding to an undetermined gravitational constant and an undetermined cosmological constant --- we know that it is at least $11$ for the gravitational theory that underlies general linear electrodynamics. Clearly, any gravitational theory with infinitely many undetermined constants is non-predictive in general, although the number of relevant constants might become finite under special circumstances, such as symmetry assumptions.

From a practical point of view, however, a general solution to the gravitational closure equations for some given matter dynamics beyond the standard model is not required. For even if it was available, it would give rise to field equations that are at least as difficult to solve as Einstein's equations. At that stage, these elusive general gravitational field equations would have to be solved either by symmetry assumptions or perturbation theory. 
The obvious idea is to implement any desired symmetry assumption or perturbative technique already at the level of the gravitational closure equations. This is not entirely straightforward, and the following two subsections briefly outline the problem an the solution.\\[6pt]
\textbf{\textit{Symmetric gravitational closure.}}
Since the gravitational closure equations yield a gravitational action, any implementation of a Killing symmetry 
$$\mathcal{L}_K G = 0$$
for the pertinent tensorial spacetime geometry at the level of the closure equations, if properly implemented, will be passed down to the action.   

But this means that variation of the action, with the aim to obtain the gravitational field equations, must now be performed with respect to the symmetric field configurations that appear in the already symmetry-reduced action. Thinking of variation in field theory as a competition between a candidate field and slightly deformed competitors, one notes that the competition is severely limited by only considering competitors that also already satisfy the imposed symmetry condition. It is thus clear that variation of a symmetry-reduced action produces weaker symmetrized field equations than variation of the full action and subsequent symmetrization would have. This is the known issue of symmetric criticality    \cite{Palais:1979} and further useful necessary and sufficient conditions for when a symmetry reduction at the level of the action yields the correct field equations have been identified    \cite{Fels:2001,Torre:2010}. 

It is clear from the above discussion that the implementation of spacetime symmetries at the level of the gravitational closure equations is possible under precisely the same conditions that apply to their implementation at the level of an action. Symmetric gravitational closure is discussed and illustrated in    \cite{DFSS}, where it is shown that one obtains the Friedmann equations as the cosmologically symmetry-reduced gravitational closure of standard model matter dynamics, without ever having to know Einstein's equations. 

For matter models beyond the standard model, the simplification of the corresponding gravitational closure equations are tremendous   \cite{DuellPhD} and thus put a solution in reach, see the contribution by D\"ull to these Proceedings. \\[6pt]
\textbf{\textit{Perturbative gravitational closure.}} 
Perturbative solutions of gravitational closure equations my be performed under precisely the same assumptions that render them meaningful at the level of the equations of motion. Since the action needs to be known, roughly speaking, to second order if one wishes to derive field equations that are valid to first order perturbations, care needs to be taken in the determination of terms that can be dropped or not. In practice, this is based on a subtle interplay of the order to which the coefficient functions of the gravitational closure equations must be expanded, the number of derivatives that act afterwards and the resulting overall order of a particular closure equation.

The perturbative treatment yields the $11$-parameter weak gravitational field equations that underlie general linear electrodynamics    \cite{Schneider} as well as the interesting bimetric gravity theory as it underlies the only superficially trivial matter model of two Klein-Gordon fields that couple to two different Lorentzian metrics, see the contribution by Wierzba to these Proceedings.

\section{Applications}
\noindent\textbf{\textit{Birefringence in the weak gravitational field of a point mass.}}
The perturbative gravitational closure of general linear electrodynamics, which served as our opening example for a problem that was previously not solvable, yields a spacetime geometry 
$$G^{abcd} = 2\eta^{a[c}\eta^{d]b} -\epsilon^{abcd} + H^{abcd}$$ 
that describes the gravitational field not too close to a point mass $m$ with the perturbative deviation $H^{abcd}$ from Minkowski space $H^{abcd}$ given in    \cite{Schneider}. No birefringence is seen, to first order perturbation theory, where and only where    \cite{Kostelecky:2001mb}
\begin{equation}
 \frac{1}{2}H^{ab}H^{cd} = \eta_{mr}\eta_{ns}H^{ambn} H^{crds} 
\end{equation}
holds to second order. Thus allowing for birefringence in principle, the weak gravitational field generated by a point mass indeed generates birefringence, whose strength depends on the mass and four more independent constants to be determined by only four experiments in that gravitational field; see also the contribution by Stritzelberger to these Proceedings. \\

\textbf{\textit{Gravitational effects in birefringent quantum electrodynamics.}}
The quantization of birefringent electrodynamics is renormalizable in a gauge-invariant way to all perturbative orders on a flat background    \cite{Grosse-Holz:2017}. This result can be used to reliably calculate quantum field theoretic processes in locally essentially flat regions of a globally non-flat area metric geometry that arises as the solution of the underyling gravity theory that one obtains by gravitational closure of classical birefringent electrodynamics. This allows to search for signatures of birefringence in localized quantum electrodynamic processes, which are now dependent of the spacetime region where they occur. Of particular interest is a modification of the observationally important $21.1$ cm line of hydrogen, which in the presence of birefringence is seen to depend qualitatively and quantitatively -- in a way precisely predicted by constructive gravity -- on the location of the hydrogen; see the contribution by Tanzi to these Proceedings.\\[6pt]
\textbf{\textit{Gravitational radiation.}}
An notable result    \cite{MoellerMSc} concerning the gravitational closure of birefringent electrodynamics is that the gravitational radiation emitted by two circularly orbiting masses contains only one massless trace-free tensor mode, as in general relativity, while additional scalar, vector and tensor modes are all massive. The production of these massive waves is shown to be significantly suppressed, since it requires the orbital frequency to surpass a certain threshold. Slowly orbiting binaries therefore only radiate waves of the type predicted by general relativity.\\[6pt]
\textbf{\textit{Etherington distance relation.}}
On a Lorentzian spacetime, the Etherington distance duality relation    \cite{Etherington} connects the luminosity distance, angular diameter distance and
redshift of an astrophysical light source independent of the gravitational dynamics. This is not the case for the refined spacetime geometry that underlies birefringent linear electrodynamics. Direct calculation, from the field equations obtained by gravitational closure, yields a modification of Etherington's relation that depends on the gravitational dynamics and indeed the particular spacetime solution    \cite{Schuller:2017dfj}. This opens up the possibility of deriving new gravitational lensing effects or indeed a pathway for the explanation of known anomalies that is directly connected to carefully studied extensions of the standard model     \cite{SME}. The reader is referred to the contribution by Werner to these Proceedings.\\[6pt]
\textbf{\textit{Parametrodynamics.}}
Parameters in matter field dynamics are usually considered as constants whose values must be determined by experiment. Gravitational closure, however, can be used to predict the values of non-scalar parameters that appear in any local field theory that is amenable to the closure equations    \cite{WierzbaMSc}. To this end, one first promotes the constant parameters to fields, analogous to the promotion of the flat spacetime Minkowski metric in Maxwell theory to the metric tensor field in the general theory of relativity. Gravitational closure of the this modified matter action then yields a multi-parameter family of actions for these parameters (and, if one so chooses, also for the underlying geometry) as the unique dynamics that enjoys a consistent co-evolution with the matter fields of the initially stipulated matter theory. See the contribution by Wierzba to these Proceedings.

\section{Conclusions}
Constructive gravity is a method to determine, by calculation rather than stipulation, a family of gravitational actions that are compatible with a large class of matter field theories. Remarkably, it is any concrete representative of these matter field theories themselves that provides the relevant information for the calculation of the gravitational theory, essentially based on the requirement that the latter have a diffeomorphism gauge group and possess a canonical evolution that shares its initial data surfaces with those of the chosen matter theory. The availability of such a procedure allows to ask and answer questions that could not be posed before and reveal gravitational dynamics as a mere consistency requirement once the matter contents of the universe is specified. While for a universe filled with standard model matter, there is no new physics predicted by the gravitational closure mechanism, this significantly changes once matter beyond the standard model is considered. This is where constructive gravity will likely find its key application. 

\section*{Acknowledgments}
The author thanks Marcus Werner for organizing a wonderful session on constructive gravity at the 15th Marcel Grossmann Meeting in Rome and for the invitation to contribute this rapporteur article to the Proceedings. He also thanks all contributors to this session for their presentations and discussions on their take of the subject and particularly Maximilian D\"ull, Florian Wolz, Nadine Stritzelberger, Alexander Wierzba, Jonas Schneider, Moritz M\"oller, Nils Fischer, Nils Alex and Hans-Martin Rieser who have contributed to the results mentioned in this article or based further work on them.

\end{document}